\documentclass[conference]{IEEEtran}
\IEEEoverridecommandlockouts
\usepackage[utf8]{inputenc}
\usepackage[T1]{fontenc}
\usepackage{cite}
\usepackage{graphicx}
\usepackage[table]{xcolor}
\usepackage{float}
\usepackage{amsthm}
\usepackage[export]{adjustbox}
\usepackage{changepage}
\usepackage{url}
\usepackage{amssymb}
\usepackage{array}
\usepackage{booktabs}
\usepackage[table]{xcolor}
\usepackage{hyperref}
\usepackage{verbatim}
\usepackage{rotating}
\usepackage{scrextend}
\usepackage{array,booktabs}
\usepackage[most]{tcolorbox}
\usepackage{dirtytalk}
\usepackage{pdfpages}

\usepackage{dblfloatfix}

\def\BibTeX{{\rm B\kern-.05em{\sc i\kern-.025em b}\kern-.08em
    T\kern-.1667em\lower.7ex\hbox{E}\kern-.125emX}}

\ifCLASSINFOpdf
\else
\fi
\begin{document}

\title{Pandemic Software Development: The Student Experiences from Developing a COVID-19 Information Dashboard
}

\author{\IEEEauthorblockN{Benjamin Koh\IEEEauthorrefmark{0},
Mojtaba Shahin\IEEEauthorrefmark{0},
Annette Ong\IEEEauthorrefmark{0}, 
Soo Ying Yeap\IEEEauthorrefmark{0},
Priyanka Saxena\IEEEauthorrefmark{0},
Manvendra Singh\IEEEauthorrefmark{0},
Chunyang Chen\IEEEauthorrefmark{0}}
\IEEEauthorblockA{\IEEEauthorrefmark{0} Faculty of IT, Monash University, Melbourne, Australia
\\ bkoh0003@student.monash.edu, mojtaba.shahin@monash.edu, \{shong21, syea0001, psax0001, msin0017\}@student.monash.edu,\\ chunyang.chen@monash.edu
}
}

\maketitle
\begin{abstract} The COVID-19 pandemic has birthed a wealth of information through many publicly accessible sources, such as news outlets and social media. However, gathering and understanding the content can be difficult due to inaccuracies or inconsistencies between the different sources. To alleviate this challenge in Australia, a team of 48 student volunteers developed an open-source COVID-19 information dashboard to provide accurate, reliable, and real-time COVID-19 information for Australians. The students developed this software while working under legislative restrictions that required social isolation. The goal of this study is to characterize the experiences of the students throughout the project. We conducted an online survey completed by 39 of the volunteering students contributing to the COVID-19 dashboard project. Our results indicate that playing a positive role in the COVID-19 crisis and learning new skills and technologies were the most cited motivating factors for the students to participate in the project. While working on the project, some students struggled to maintain a work-life balance due to working from home. However, the students generally did not express strong sentiment towards general project challenges. The students expressed more strongly that data collection was a significant challenge as it was difficult to collect reliable, accurate, and up-to-date data from various government sources. The students have been able to mitigate these challenges by establishing a systematic data collection process in the team, leveraging frequent and clear communication through text, and appreciating and encouraging each other's efforts. By participating in the project, the students boosted their technical (e.g., front-end development) and non-technical (e.g., task prioritization) skills. Our study discusses several implications for students, educators, and policymakers.

\end{abstract}


\section{Introduction}\label{sec:Introdcution}
The emergence and spread of the coronavirus disease 2019 (COVID-19) have had a global impact, with 190 countries having reported cases as of April 5, 2020 \cite{johnhopkins2020covid}. In Australia, federal restrictions were enacted to reduce community transmissions, which have resulted in the closure of work premises and university campuses. Consequently, work and collaboration have needed to shift into remote and online environments. The distribution and reporting on the various aspects of COVID-19 through multiple sources, such as federal reports or social media, have posed a challenge for Australians to understand the numerous facets of the pandemic situation. In light of this, a team of 48 student volunteers developed an open-source COVID-19-AU information dashboard (\textcolor{blue}{https://covid-19-au.com}) while working under legislative restrictions that required social isolation (inspired by \cite{ralph2020pandemic}, we refer to this sort of software development as \textit{pandemic software development}). The project is a data aggregation software that aims to ease the challenge of understanding COVID-19 in Australia by providing reliable, accurate, and updated information about the pandemic. The site displays useful metrics, data visualizations, and relevant articles derived from various sources.


Some prior research on developing COVID-19 software has concentrated on technical challenges like bugs and improving software quality \cite{Rahman2020exploratory}, and characteristics of COVID-19 software repositories \cite{wang2020open}. Other researchers focus on how the pandemic has influenced educational institutions and their instructional strategies (e.g., \cite{bao2020covid, kanij2020adapting, barr2020online}). A growing number of the literature \cite{ford2020taleNew, butler2020challenges, bao2020does, bezerra2020human, ralph2020pandemic, rodeghero2020please, russo2021predictors, o2021covid, miller2021your} has started looking into the consequences of the pandemic on the software development process (e.g., the effects of relocating into a new working environment with working from home \cite{ford2020taleNew}). These works investigate the problems, motivations, and experiences of \textit{practicing professionals}. Some other researchers investigated the different aspects of engaging students in open-source software development (e.g., \cite{silva2020google, pinto2019training, smith2014selecting}). Despite these efforts, there is no research on how a \textit{student team} develops a software project during the COVID-19 pandemic.
The challenges faced by a (cross-disciplinary) \textit{student team} may vary from that of a \textit{professional team} due to various factors like lack of prior work experience or the absence of pre-existing policies to aid work management during a crisis. Hence, these differences may display varying behavior in how the students adapt and overcome these challenges.


The research presented in this paper aims to comprehend the experiences of the students that contributed to the COVID-19-AU information dashboard project in four areas: their motivations to contribute to the project, the challenges while working on this project, the strategies employed to overcome the challenges, and the learning's gained through the experiences. To this end, we conducted an online survey with 39 out of 48 project participating students, with questions formulated based on a pilot analysis of the student team's internal messages on Slack and their usage of GitHub, as well as existing literature.
\begin{figure*}[t]
\centering
\includegraphics[scale=0.4]{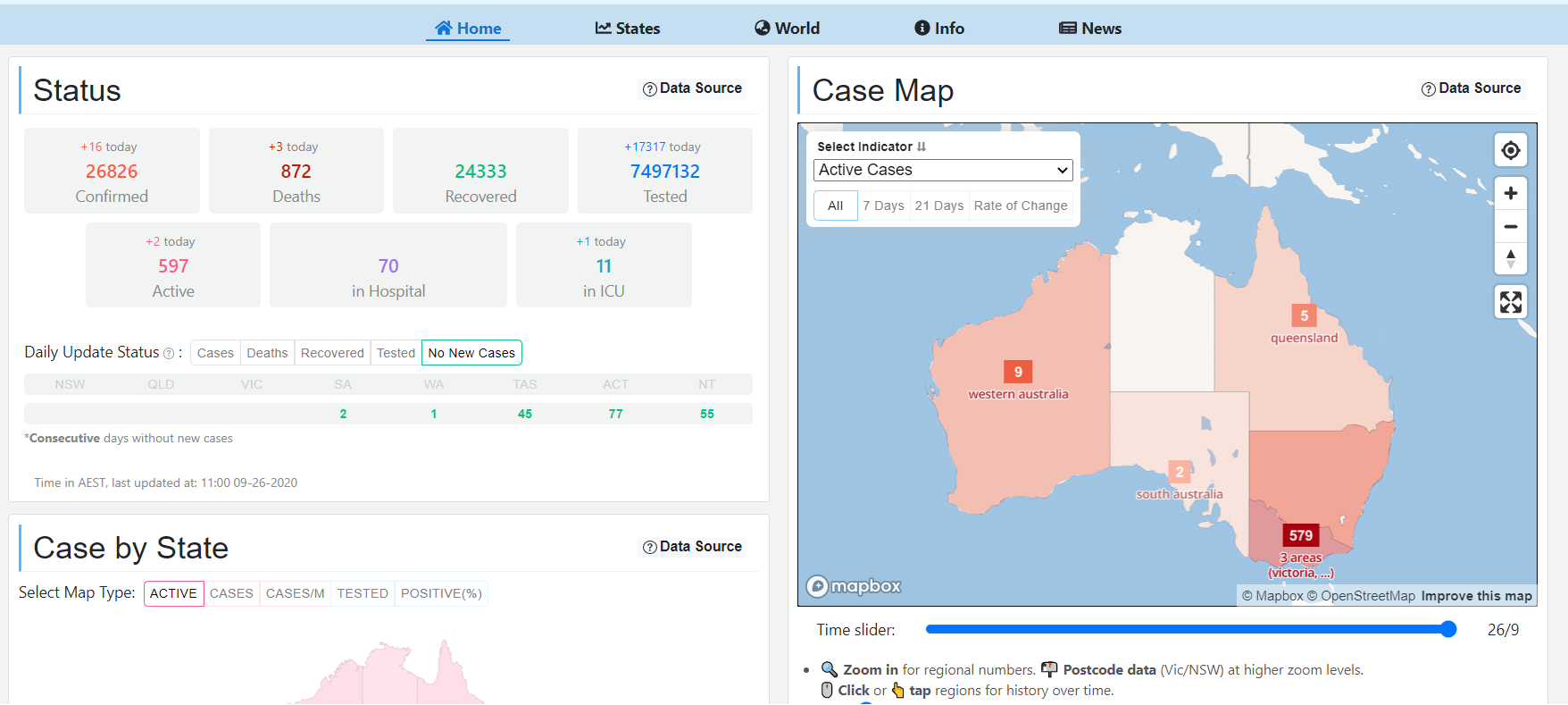}
\caption{A Snapshot of the COVID-19 Information Dashboard (https://covid-19-au.com)}
\label{fig:dashboard}
\end{figure*}

Our findings indicate that playing a positive role in the crisis was the motivation most selected by the participants. Other significant motivations include learning new skills and technologies and gaining experience for their CV. The most prominent challenges for the students were regarding data collection, which was strongly attributed to the unreliability and inaccuracy of the data provided through government sources. Some students expressed difficulty in maintaining a work-life balance while working from home and in forming relationships with the other contributors as the project's development was entirely remote. However, the majority of general project challenges were not deemed substantial by the students, which ties with the many mitigation strategies that were employed by the students for the duration of the project. Of these mitigation strategies, the students most frequently used clear language when messaging other team members. Team morale was also maintained through vocal appreciation of the work and by encouraging the other contributors. Overall, the students exhibited growth in their technical and non-technical skills, particularly in their front-end development skills, understanding of COVID-19, and ability to work online.

\textbf{Contributions}: (1) a better understating of the student experience in pandemic software development (2) a set of practical implications for students, researchers, and policymakers.

\textbf{Paper Structure}: Section \ref{sec:Background} provides the background. Section \ref{sec:ResearchMethod} explains our research method. Section \ref{sec:Findings} reports the findings. We reflect on our findings in Section \ref{sec:Discussion}. Section \ref{sec:RelatedWork} summarises the related work. Section \ref{sec:Conclusion} concludes our study. 
\section{Background}\label{sec:Background} 
\subsection{COVID-19}
As of February 5, 2021, there have been 130, 422,190 confirmed cases of COVID-19 and 2, 842,135 confirmed deaths \cite{whodashboard}. Like other countries, the Australian government introduced legislative restrictions to minimize the spread of the disease \cite{ausgovnews}. 
These restrictions included limiting public gatherings, shutting down on-site working alongside encouraging working from home (WFH), placing a 5km radius restriction on an individual traveling from their house, and limiting the reasons for leaving the house. These restrictions resulted in the closure of Australian university campuses, with students and staff needing to transition into an online environment \cite{anneausuni}.

\subsection{COVID-19 Information Dashboard}
Through the progression of this pandemic, a plethora of information about the disease has become publicly available through many different sources, such as news outlets, federal and state reports, and social media \cite{storenreporting}. The consolidation of these sources can be overwhelming due to inaccuracies or inconsistencies in their content. The COVID-19 information dashboard was developed as a data aggregation software for COVID-19 information in response to this challenge. The website aims to help Australians stay informed on key areas of the disease, such as case numbers and outbreak hot spots. Fig. \ref{fig:dashboard} is a snippet of the dashboard.

\subsubsection{Contributors}
The COVID-19 information dashboard project is open-source software (OSS) maintained and managed by a team of 48 student volunteers. The students were recruited primarily through advertisements on student social spaces and by word of mouth. The team includes four key areas: data collection, data visualization, web development, and marketing. Many of the students did not have prior expertise in relation to their area of contribution. The majority of the students also had no prior professional experience in working on a software project (74.36\%). A further breakdown of the team's demographics is presented in Section \ref{sec:demographics}.

\subsubsection{Features}
The COVID-19 information dashboard displays a variety of metrics on COVID-19. Metrics including confirmed cases, active cases, recovered cases, test numbers, hospital numbers, and deaths are presented both numerically and through heat map visualizations. The specificity of the data can also be varied between a national view, state view, and local government area view. In addition, time series data visualizations are also available to view historical figures or trends in the data. Other key features of the website also include a global comparison of COVID-19 between Australia and other nations, demographic analysis of the pandemic by state, and a timeline of COVID-19 related news articles released by reputable outlets. The data used by the site is updated in real time through the use of an automated data crawler that scrapes various sources, such as government reports, news articles, and social media posts. The data also undergoes a manual review to manage inconsistencies or mistakes. 

\subsubsection{Development Process} \label{sec:developmentprocess}
Due to the restrictions in Australia, the team was required to collaborate through a purely online medium, with many also having to contribute while working from home. Communication within the team was largely facilitated through the use of Slack, which provides private messaging, channel messaging, and video conferencing services. In addition, GitHub was used as the version control hosting site for the software. The developers would perform issue labeling and provide feedback through the features provided by the platform. Moreover, they had to self-manage and develop the necessary policies and processes for project development. This included the on-boarding process of new members, managing shared resources such as spreadsheets and code, and the workflows required to ensure ongoing tasks could be tracked. Many strategies adopted by the team in developing the software are discussed in this paper.

\subsubsection{Project Impact}
The project has gathered much public attention, with the website achieving almost 13 million visits as of April 5, 2020. Among all the visitors, more than 200,000 (22.2\%) users are returning users.
It has attracted over 700,000 unique users since its release and had almost 60,000 users access the site in a single day during the peak of the pandemic in Australia. The site has also attracted visitors from 180 other countries around the world.
There have been 3,641,690 sessions, with users spending 2.62 minutes on average in each session. User feedback collected through different channels (e.g., social media, email, online survey) showed that around 81\% of the users had positive reviews of the website. Examples of positive user sentiments include \emph{``it is so impressive to see the additional features that seem to be added daily''} and \emph{``I check every day for updates. I've shared with many friends and family, and we all think you've done an amazing job''}. Feature requests from users were also used in developing features on the site.


\section{Research Method}\label{sec:ResearchMethod}
To characterize the experiences of the student volunteers of the COVID-19 dashboard project (for brevity, the COVID-19 project) during the pandemic, we conducted a survey. 
\subsection{Research Questions}
Our study was guided by the following research questions:
\begin{center}
\begin{tcolorbox}[colback=green!2!white,colframe=black!75!black]
\textit{\textbf{RQ1.} What factors do motivate students to contribute to the COVID-19 project during the COVID-19 pandemic?}
\end{tcolorbox}
\end{center}
\textbf{\underline{Motivation}}: Students may have participated in the COVID-19 project for different reasons, such as improving their CVs and challenging themselves. This question aims to understand what factors motivated students to work on the COVID-19 project during the COVID-19 pandemic.

\begin{tcolorbox}[colback=green!2!white,colframe=black!75!black]
\textit{\textbf{RQ2.} What challenges do students experience when working on the COVID-19 project during the COVID-19 pandemic?}
\end{tcolorbox}

\textbf{\underline{Motivation}}: Software development is a collaborative and joint endeavor that involves individuals with diverse skills and seniority. The COVID-19 pandemic may negatively influence this process. For example, participating students may have had a lack of or inadequate access to (accurate) resources (e.g., data). They may have faced new or exacerbated software bugs that they did not experience before \cite{Rahman2020exploratory}. Hence, we aim to understand such challenges that may have impacted students’ learning process and work on the COVID-19 project.
\begin{tcolorbox}[colback=green!2!white,colframe=black!75!black]
\textit{\textbf{RQ3.} What strategies do students employ when working on the COVID-19 project during the COVID-19 pandemic?}
\end{tcolorbox}

\textbf{\underline{Motivation}}: It aims to understand the strategies that students consider beneficial in minimizing the impact of the pandemic on the project and their involvement in the project.
\begin{tcolorbox}[colback=green!2!white,colframe=black!75!black]
\textit{\textbf{RQ4.} What skills do students learn when working on the COVID-19 project during the COVID-19 pandemic?}
\end{tcolorbox}


\textbf{\underline{Motivation}}: Participating students may have learned and developed new technical and non-technical skills and expertise while working on the project within the COVID-19 pandemic. Such skills and experiences may be leveraged by students and their peers in the future to deal with such kind of crises and disasters in their personal life or professional career.

\subsection{Pilot Study}
Research on the effects of the COVID-19 pandemic on software development and software engineering education is rare. Hence, it was challenging for us to design appropriate and comprehensive survey questions. To better understand the possible challenges faced by and solutions adopted by students, we conducted a pilot study before the survey to analyze the team discussions on Slack and the issue tracking system of the project. 

\begin{table}[]
\caption{Top 10 challenges highlighted by Slack messages}
\label{tba:SlackMesage}
\centering
\resizebox{.47\textwidth}{!}{%
{\renewcommand{\arraystretch}{1.0}
\begin{tabular}{p{6cm} c}
\hline
\textbf{Code}                                   & \textbf{Total} \\ \hline
Collecting data                                 & 147            \\
Designing the interface                         & 57             \\
Working with the codebase                       & 57             \\
Having the necessary experience for the project & 29             \\
Time management with project tasks              & 29             \\
Mistakes in day entry                           & 22             \\
Online communication                            & 15             \\
Heavy workload                                  & 11             \\
Using internal team resources                   & 11             \\
Balancing external commitments                  & 9              \\ \hline
\end{tabular}

}\quad
}
\end{table}

\textbf{Slack}. First, the text dump of messages was cleaned, and non-text blocks such as links and code snippets were removed. Next, the text was tokenized, and the tokens were categorized by research question. The tokens were then used as search terms for relevant messages, with further filtering was done to remove false positives. Finally, the relevant messages were open coded \cite{creswell2002educational} into categories that would form the basis of our survey questions. For instance, the challenge of \emph{collecting data} made up a high proportion of challenge-related messages (147 out of 457, or 32.17\%), as shown in Table \ref{tba:SlackMesage}. This informed several data collection specific challenges that were presented in a separate section of the survey. Additionally, we also observed a substantial proportion of messages centered around proactive collaboration, such as offering to help with debugging or acknowledging team members' work. As a result, \textit{``actively responding to teammates' messages''} and \textit{``encouraging the efforts of other members''} were included as strategy statements in the survey.

\textbf{GitHub}. We also analyzed comments in pull requests and raised issues to identify common topics discussed among the participants. Of 66 discussion threads, most of them centered around \textit{bug fixes} and \textit{code review}, further reinforcing the Slack data analysis that showed participants lacked technical experience and working with the codebase. Another discussion details how participants worked together to establish a reliable source for the number of recovered cases to replace the Australian government website when it stopped releasing them in late March. This resulted in providing the basis for the strategy \textit{``prompt discussing any differences found across data sources with my team''} to be included in the survey.

\subsection{Main Study: Survey}
\subsubsection{Protocol}
We designed an online survey that included both close-ended and open-ended questions. The survey was hosted on Qualtrics software\footnote{https://www.qualtrics.com/au/core-xm/survey-software/} and took approximately 25 minutes to complete. The complete list of the questions is available at \cite{replpack}.
The survey questions were mainly formulated based on the existing literature (e.g., \cite{ford2020taleNew, da2010challenges,silva2020google}) and our pilot study (i.e., analyzing the students' discussions on Slack and GitHub). The survey started with a brief overview of our research objective. The survey questions can be generally organized into six groups:

    \textbf{Demographics}. Our survey had 7 questions to collect demographic information such as students' gender and their prior professional experience in software development.
    
    \textbf{Motivations}. Inspired by \cite{silva2020google, lakhani2003hackers}, we designed a multiple-choice question to seek the students' motives to contribute to the project. 
    
    \textbf{Challenges}. We used 10 statements (i.e., some of them adopted from \cite{ford2020taleNew, da2010challenges}) to measure to what extent students agreed or disagreed with the potential challenges (if any) that may have occurred during their work on the COVID-19 project. The statements were rated based on a six-point Likert scale (from ``Strongly agree'' to ``Strongly disagree''). We also provided “Not Applicable” in the answer options, which allowed students to indicate if they did not face a given challenge. We also designed 4 six-point Likert scale statements to seek the level of agreement or disagreement of students on the possible challenges in data collection during the project.
    
    \textbf{Practices}. Similarly to the challenges, we designed two types of statements to explore how often students applied a provided list of 10 practices or techniques to address the challenges they encountered during the project, as well as 5 used exclusively during the data collection. The students were asked to rate them using five-point Likert scale statements (from ``Very often'' to ``Never'').       
    
    \textbf{Skills}. The students were asked to show how confident they were about 8 soft skills and 8 technical skills before contributing to the project and after participating in the project. These statements were rated on 11 scales from 0 (not confident) to 10 (very confident).
    
   \textbf{Open-ended Questions}. All multiple- and single-choices questions (except for the demographic questions) were followed by an open-ended question to collect further opinions and thoughts from the students. In total, we asked 7 open-ended questions.


\subsubsection{Participants}
The survey participants were recruited from the volunteering students who contributed to the COVID-19 information dashboard. In total, 48 students were involved in different steps of the project. After collecting the email addresses of the 48 students, we sent an invitation email to all of them. We also sent a message via Slack and invited them to fill in the survey. To encourage students to complete the survey, we promised the students that they would receive an incentive of an 80 AUD Coles gift card for participating. We received 39 responses (acceptance rate: 81.25\%).


\subsection{Data Analysis}
Closed-ended questions were analyzed using descriptive statistics, while open-ended questions were analyzed using open coding \cite{creswell2002educational} and axial coding \cite{corbin2014basics}. The codes generated were used to provide more descriptive results for the related closed questions.

\section{Findings}\label{sec:Findings}
\subsection{Demographics}\label{sec:demographics}
\textbf{Age}. The participants' ages ranged from 20 years old to 35 years old. The mean age of the participants is 24.56 years.

\textbf{Gender}. Of 39 participants, 27 (69.23\%) were male, and the rest (12, 30.77\%) were female.

\textbf{Education}. 56.41\% of the participants stated that they were doing a postgraduate degree, and 43.59\% said that they were pursuing an undergraduate degree at the time of the survey. 

\textbf{Degree}. The majority of the students' degrees (69.23\%) were related to IT fields such as Information Technology, Software Engineering, Data Science, and Computer Science. The rest (30.76\%) were studying non-IT-related disciplines, e.g., Education, Mechanical Engineering, Law, and Art.

\textbf{Role}. 35.90\% of the students indicated that they mainly contributed to the project as a developer. 12 students (30.77\%) were primarily responsible for marketing. Information contributor was another highly cited role by the students (23.08\%).

\textbf{Area of Contribution}. Table \ref{table:team_breakdown} depicts the different aspects of the project and the proportion of survey participants identified as having worked in each area. A single team member could select multiple areas. The most dominant areas of the project were data collection (17, 43.59\%) and promotion/marketing (17, 43.59\%).

\begin{table}[]
\caption{Team Breakdown and Responsibilities}
\label{table:team_breakdown}
\centering
\resizebox{.47\textwidth}{!}{%
{\renewcommand{\arraystretch}{1.01}
\begin{tabular}{l l l}
\hline
\textbf{Team} & \textbf{\#Num} & \textbf{Responsibility}  \\ \hline
Promotion & 17 &  Promote the website in different channels \\  
Data collection & 17 & Collect general info (e.g., policy, symptom)   \\                               
Data update & 12 & Live number update from government site\\
Front-end development & 11 & Develop front-end website \\
Data vis & 10 & Visualise the statistics interactively\\
UI/UX & 8 & Graphical UI design and user experience \\
Back-end development & 4 & Develop the back-end \\
Translation & 3 & Develop multi-lingual site\\ 
\hline
\end{tabular}
}\quad
}
\end{table}

\textbf{Experience}. 29 students (74.36\%) pointed out that they had no experience in software development in the industry before joining the project. Only 10 students (25.64\%) indicated that they had such experience: 4 had 1-2 years experience, 5 had 3-5 years experience, and 1 had more than 10 years experience.

\subsection{RQ1: Motivating Factors}
The participants were presented with 12 motivations and asked to select the statements that applied to them. Table \ref{table:Motivations} shows the motivations that the participants selected from.



\begin{table}[]
\centering
\caption{List of motivating factors and survey responses}
\resizebox{\linewidth}{!}{
{\renewcommand{\arraystretch}{1.01}
\begin{tabular}{lcc}
\hline
\textbf{Statements} & \textbf{\#}  & \textbf{\%}   \\ \hline
Play a positive role in this crisis & 35 & 89.74 \\
Learn new skills and technologies & 30 & 76.92 \\ 
Gain experience and build CV & 27 & 69.23 \\ 
Build a network with new people & 26 & 66.67 \\ 
Participate in a student initiated/led project  & 25 & 64.10 \\ 
Learn more about the COVID-19 situation & 24 & 61.54 \\ 
Demonstrate my skills and expertise & 23 & 58.97 \\ 
Challenge myself to do technically challenging tasks & 19 & 48.72 \\ 
Experience working on a crisis related software project & 19 & 48.72 \\ 
Learn to develop software system in industrial setting & 17 & 43.59 \\ 
To occupy free time & 10 & 25.64 \\ 
Others & 3 & 7.69 \\ \hline
\end{tabular}
}
}\quad

\label{table:Motivations}
\end{table}

Given the ongoing pandemic, 89.74\% of participants were highly motivated to play a positive role in the crisis. 58.97\% took the project as an opportunity to demonstrate their skills, with one participant commenting he could use his past experience in \emph{``projects (that...) involved large-scale data collection and normalization''} to contribute \emph{``towards Australian and international efforts combating this pandemic''}. 61.54\% wanted to learn more about COVID-19, and 48.72\% also saw this as an opportunity to experience working on a crisis related software project. 

As mentioned earlier, 74\% of the students had no software development experience in the industry before joining the project, which may have resulted in the team being strongly motivated by the desire to learn new skills and technologies (77\%), and gaining experience for their CV (69\%). Moreover,  participants further expressed they wanted to build a network by meeting and communicating with new people (67\%) and to contribute to a student-initiated/led project (64\%). 
Participants largely disagreed with participating simply to occupy free time, with only 26\% agreeing with the option. 


\begin{center}
\begin{tcolorbox}[colback=black!5!white,colframe=black!75!black]
\textbf{\footnotesize RQ1.} \textit{We found that playing a positive role in the COVID-19 crisis, learning new skills and technologies, and gaining experience for their CV were the most cited motivations for volunteering students to contribute to the COVID-19 information dashboard.}
\end{tcolorbox}
\end{center}

\subsection{RQ2: Challenges} \label{sec:Challenges}
We classify the possible challenges that the students faced into General Challenges and Data Collection Challenges. 

\begin{table*}[!ht]
\centering
\caption{List of general challenges and responses (in \%). STA: Strongly Agree, A: Agree, SWA: Somewhat Agree, SWD: Somewhat Disagree, D: Disagree, STD: Strongly Disagree}
\footnotesize
\resizebox{\textwidth}{!}{%
{\renewcommand{\arraystretch}{1.0}
\rowcolors{2}{gray!25}{white}
\begin{tabular}{|c|l|cccccc|}
\hline
  \textbf{ID}  &                                                                              \textbf{Challenge}                 & \textbf{STA}  & \textbf{A}     & \textbf{SWA}   & \textbf{SWD}   & \textbf{D}     & \textbf{STD}   \\ \hline
C1  & Balance my university commitments and working on this project    & 2.94 & 8.82  & 26.47 & 14.71 & \textbf{32.35} & 14.71 \\ \hline
C2  & Manage my mental health while working on this project from home         & 0.00 & 5.26  & 7.89  & 7.89  & 31.58 & \textbf{47.37} \\ \hline
C3  & Manage physical health while working on this   project                 & 2.56 & 2.56  & 2.56  & 7.69  & \textbf{43.59} & 41.03 \\ \hline
C4  & Work on this project from home                                     & 0.00 & 7.69  & 5.13  & 12.82 & 33.33 & \textbf{41.03} \\ \hline
C5  & Keep attention away from distractions while working from home & 0.00 & 17.95 & 15.38 & 12.82 & \textbf{33.33} & 20.51 \\ \hline
C6  & Establish a work-life balance while working from home                     & 5.13 & 20.51 & \textbf{23.08} & 7.69  & 20.51 & \textbf{23.08} \\ \hline
C7  & Set up my own productive workspace while working from   home              & 2.56 & 12.82 & 10.26 & 10.26 & \textbf{33.33} & 30.77 \\ \hline
C8  & Stay motivated while working from home                             & 5.13 & 12.82 & 15.38 & 20.51 & \textbf{25.64} & 20.51 \\ \hline
C9 & Collaborate with the team through a purely   online medium         & 2.56 & 12.82 & 12.82 & 17.95 & \textbf{28.21} & 25.64 \\ \hline
C10 & Work in a purely online medium                                     & 0.00 & 7.89  & 5.26  & 18.42 & \textbf{44.74} & 23.68 \\ \hline
\end{tabular}
}\quad
}
\label{table:challenges}
\end{table*}

\begin{table*}[!ht]
\centering
\caption{List of data collection challenges and survey responses (in \%). STA: Strongly Agree, A: Agree, SWA: Somewhat Agree, SWD: Somewhat Disagree, D: Disagree}
\resizebox{\textwidth}{!}{%
{\renewcommand{\arraystretch}{1.0}
\rowcolors{2}{gray!25}{white}
\begin{tabular}{|c|p{9cm}|cccccc|}
\hline
\textbf{ID}  &                                                                  \textbf{Statements}         & \textbf{STA} & \textbf{A} & \textbf{SWA} & \textbf{SWD} & \textbf{D} & \textbf{STD} \\ \hline
C11 & Find reliable and accurate data                       & 0.00           & \textbf{47.06} & 29.41          & 5.88              & 17.65    & 0.00              \\ \hline
C12 & Find data that was up to date                         & 5.88           & 17.65 & \textbf{52.94}          & 5.88              & 17.65    & 0.00              \\ \hline
C13 & Find data that was consistent across multiple sources & 11.76          & \textbf{35.29} & 23.53          & 17.65             & 11.76    & 0.00              \\ \hline
C14 & Interpret the data on sources we found                & 0.00           & \textbf{35.29} & 5.88           & 23.53             & 29.41    & 5.88              \\ \hline
\end{tabular}
}\quad
}
\label{table:data_challenges}
\end{table*}

\subsubsection{General Challenges}
We showed the students a list of possible challenges (See Table \ref{table:challenges}). None of the challenges were agreed upon by more than 50\% of the participants. Furthermore, \emph{Strongly Disagree} and \emph{Disagree} were the most selected options for this question. Our study also showed that this sentiment was not sensitive to the participant's background, as heavy disagreement was also observed when breaking down the results by demographics such as prior experience and role. While these challenges were present in the pilot study, the students did not feel strongly about their presence in the project. This relates to the active employment of several strategies, which is further discussed in Section \ref{sec:strategies}. Notably, the participants did not feel that it was challenging to manage their mental and physical health when working on the project during the pandemic (statements C2 and C3 respectively), as over 80\% disagreed with these statements. This contrasts previous research that reports an increase in psychological distress levels during the pandemic \cite{anuhardship}.

The vast majority of the participants (86.84\%) did not agree that working only through an online medium was a challenge (statement C10), and 71.8\% of them felt that it was not difficult to collaborate through a purely online medium (statement C9). Despite this, we found some references in the open-ended questions, which indicates that team building and relationship forming was difficult due to having to communicate through a purely online medium. For example, a student mentioned that \emph{``the sense of belonging and community was lacking a bit in my opinion''}, and another wrote \emph{``having never met any of the other student volunteers in real life (...) was a little challenging''}. Other students described the effect of such a working style on team and project management. As an example, we have, \emph{``It was also challenging to manage the team, especially when delegating tasks because many people would join the project but may not have real commitment''}. Of the general challenges, the most prevalent was 50\% of students agreeing that establishing a work-life balance while working from home was a challenge (statement C6), which is related to the government 'stay-at-home' policies and its impact on students in managing the different areas of their lives.

\begin{table*}[!ht]
\centering
\caption{List of general strategies and practices and survey responses (in \%)}
\resizebox{\textwidth}{!}{%
{\renewcommand{\arraystretch}{1.0}
\rowcolors{2}{gray!25}{white}
\begin{tabular}{|c|p{10cm}|ccccc|}
\hline
\textbf{ID}   &                                                                          \textbf{Statements}                 & \textbf{Never} & \textbf{Rarely} & \textbf{Sometimes} & \textbf{Often} & \textbf{Very} \textbf{Often} \\ \hline
S1 & 
Worked on tasks with teammates & 2.56  & 10.26  & \textbf{38.46}     & 35.90 & 12.82      \\ \hline
S2 & 
Worked according to fixed schedule & 2.56  & 5.13   & 20.51     & \textbf{48.72} & 23.08      \\ \hline
S3 & 
Used clear language in messages & 0.00  & 0.00   & 7.69      & 35.90 & \textbf{56.41}      \\ \hline
S4 & 
Used voice calls for discussions & 15.38 & \textbf{35.90} & 25.64 & 10.26 & 12.82 \\ \hline
S5 & 
Responded to teammates' messages actively & 0.00  & 0.00   & 12.82     & \textbf{46.15} & 41.03      \\ \hline
S6 & 
Encouraged and acknowledged the efforts of other teammates & 0.00  & 7.69   & 15.38     & \textbf{38.46} & \textbf{38.46}      \\ \hline
S7 & 
Celebrated project milestones with the team & 5.13  & 5.13   & 28.21     & \textbf{38.46} & 23.08      \\ \hline
S8  & 
Distanced self from constant feed of COVID-19 related news & 17.95 & 25.64 & \textbf{30.77} & 17.95 & 7.69  \\ \hline
S9 & 
Referred to other COVID-19 projects to help create visualisations & 20.51 & 10.26 & 25.64 & \textbf{28.21} & 15.38 \\ \hline
S10 & 
Set up workspace separate from personal space & 12.82 & 17.95 & 23.08 & \textbf{28.21} & 17.95 \\ \hline
\end{tabular}
}\quad
}
\label{table:generalStrategies}
\end{table*}

\begin{table*}[!ht]
\centering
\caption{List of data collection strategies and practices and survey responses (in \%)}
\resizebox{\textwidth}{!}{%
{\renewcommand{\arraystretch}{1.0}
\rowcolors{2}{gray!25}{white}
\begin{tabular}{|l|p{10cm}|ccccc|}
\hline
   \textbf{ID} & \textbf{Statements}                                                                                & \textbf{Never} & \textbf{Rarely} & \textbf{Sometimes} & \textbf{Often} & \textbf{Very Often} \\ \hline
S11 & Maintained a list of trustworthy sources     & 0.00  & 0.00   & 12.00     & 41.00 & \textbf{47.00}      \\ \hline
S12 & Regularly revisited data sources to collect the most up-to-date information             & 0.00  & 0.00   & 0.00      & 41.00 & \textbf{59.00}     \\ \hline
S13 & Discussed any differences found across data sources with my teammates        & 0.00  & 6.00  & 12.00 & \textbf{47.00} & 35.00 \\ \hline
S14 & Cross-checked data with multiple sources                   & 0.00  & 0.00   & 0.00      & \textbf{59.00} & 41.00      \\ \hline
S15 & Did background research on topics related to COVID-19    & 0.00  & 6.00   & 18.00     & 35.00 & \textbf{41.00}     \\ \hline
\end{tabular}
}\quad
}
\label{table:dataStrategies}
\end{table*}

\subsubsection{Data Collection Challenges}
Data collection was one of the most dominant activities of the project, with 42.5\% of participants indicating that they were involved in data collection. Challenges related to data collection were also frequently found through the pilot study, with example Slack messages such as \emph{``gov can make a mistake and it's hard for us to always keep an eye on it''}, \emph{``they are always changing the format...''}, and \emph{``due to the lack of a federal level information hub we have to manually collect some data for our app from different places''}.  Hence, we presented the participants involved in data collection with challenge statements specific to data collection, shown in Table \ref{table:data_challenges}. We observe the majority of students agreeing to the statements, which shows that these challenges were prevalent in the project.

Moreover, analysis of the open questions also highlighted the issues associated with the government's reporting process in releasing information about COVID-19. One student commented that \emph{``Governments release their data at different times, which requires us to monitor and collect data throughout the entire day. This increased the workload of the job''}, and another commented, \emph{``state governments had disparate processes which made our jobs a bit harder''}. Thus, we attribute the challenges faced by students in collecting data to the reporting process by the government during the pandemic. Conversely, while collecting data was a challenge for many of the participants, the team did not consider understanding the data found on the sources difficult. Approximately 59\% of participants disagreed that it was difficult to interpret the data on the sources they found (statement C14).

\begin{center}
\begin{tcolorbox}[colback=black!5!white,colframe=black!75!black]
\textbf{\footnotesize RQ2.} \textit{Students largely disagreed with the general challenges, including management of their mental health. They voiced that forming relationships and team building was difficult as the project was entirely online. They expressed more strongly that data collection was a significant challenge as it was difficult to collect reliable, accurate, and up-to-date data from various government sources.}
\end{tcolorbox}
\end{center}

\subsection{RQ3: Strategies and Practices} \label{sec:strategies}

We presented a total of 15 strategy statements, of which 5 were specifically related to data collection. 
We considered a strategy to be have been commonly used if they were rated ``Often'' or ``Very Often'' by the participants.

\subsubsection{General Strategies}
As shown in Table \ref{table:generalStrategies}, the most commonly used practices were related to communication and team well-being management. Statements S3 and S5, which were related to communication, had 92\% and 87\% of the participants indicating that they were common practices. These were practices the participants have adopted to address the communication challenges discussed in Section \ref{sec:Challenges}. The main form of communication between the project participants was through text messages rather than voice calls, as only 21\% agreed that voice calls were made for more efficient conversations (statement S4). 

Efforts were made to manage individual well-being. To keep the team motivated, the participants often encouraged each other (statement S6) and celebrated project milestones as a team (statement S7). In responses to the open question, a few participants also mentioned practices to manage their well-being, such as having \emph{``regular exercise, consistent sleep schedule…''} and \emph{``…periods of relaxation…''}. Another participant mentioned \emph{``building personal relationships with team members … helped during isolation''}. About 72\% of the participants also responded that they worked on a fixed schedule to manage their commitments. The open-ended questions also revealed other practices in task management, with participants mentioning that they kept track of tasks through reminders and referred to different resources to help them with their tasks.

We identified a few strategies that the participants less commonly practiced. About 56\% of the participants rarely distanced themselves from the constant feed of COVID-19 news (statement S8), and close to 18\% of the participants never did so. This strategy could have been difficult to practice as the majority of the tasks revolved around data collection. Another practice that was not as popular was to refer to similar sites for data visualization creation (statement S9). Above 20\% never adopted this practice, and about 36\% did not do so frequently. This can be justified by the fact that only a quarter of the participants were involved in data visualizations. About half the participants also infrequently separated their work and personal spaces (statement S10), with only 46\% stating they have kept separate spaces for different purposes. 

\subsubsection{Data Collection Strategies}
Most participants frequently practiced the strategies presented in the survey, with participants agreeing that they were (very) often visiting data sources for most up-to-date data (statement S12) and cross-checking data across sources (statement S14). These practices helped the participants overcome the challenges in finding reliable and accurate data (statement C11) and up-to-date data (statement C12). Responses to the open-ended questions also supported the use of these strategies, with participants mentioning Twitter updates and referring to other sites for inspiration.

\begin{center}
\begin{tcolorbox}[colback=black!5!white,colframe=black!75!black]
\textbf{\footnotesize RQ3.} \textit{Frequent and clear communication through text was one of the most commonly adopted practices. Students also frequently showed appreciation and encouragement to other team members. As for data collection, students often visited trusted sources and cross-checked data to ensure the quality of data displayed on the website.}
\end{tcolorbox}
\end{center}

\subsection{RQ4: Technical and Soft Skills} 
In order to determine how much the participants felt they learned from their participation in the project, we presented a set of 16 skills (8 technical and 8 non-technical). Each participant was asked to rank on a numerical scale of how confident they felt in each skill (0 being the least confident and 10 being the most confident) before and after the project.

\subsubsection{Technical Skills}\label{sec:techSkills}

\begin{figure}[h]
    \centering
    \includegraphics[width=8.5cm,height=6cm,keepaspectratio]{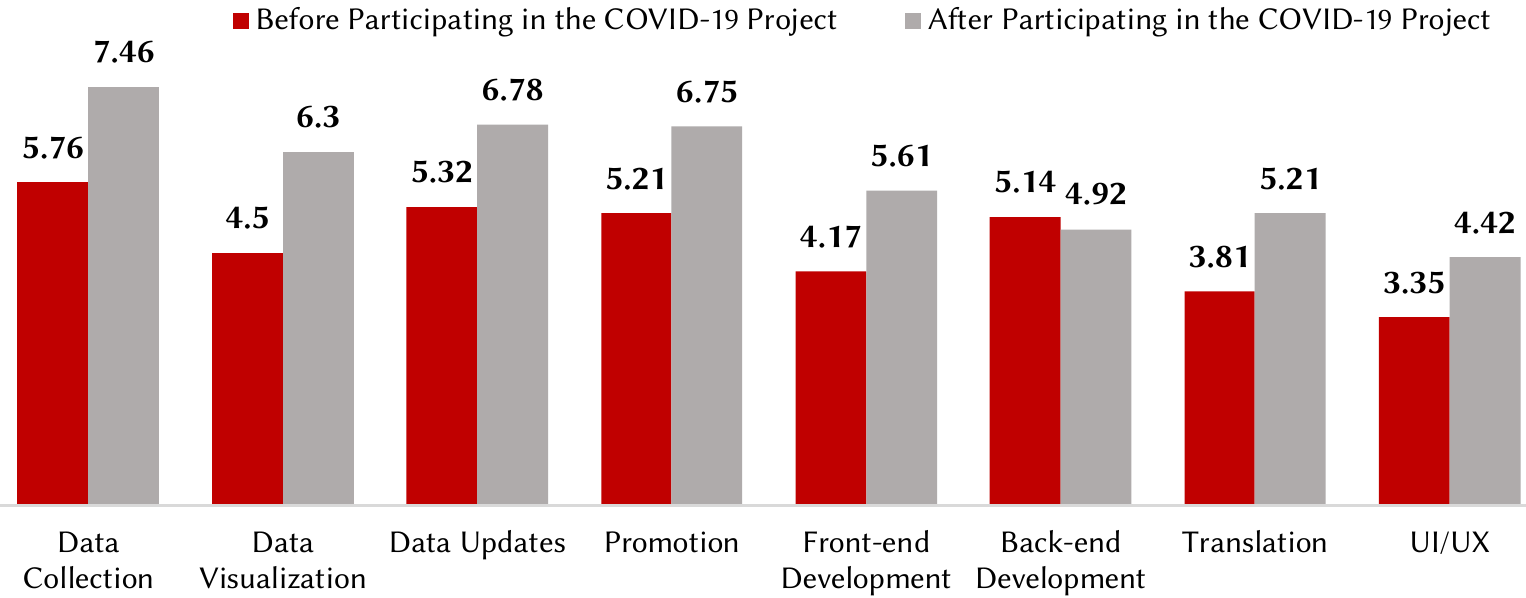}
    \caption{The mean of the participants' confidence in technical skills before and after participating in the COVID-19 project}
    \label{fig:techskills}
\end{figure}
As shown in Fig. \ref{fig:techskills}, out of software development-related skills, participants grew the most in front-end development with 34\%, while back-end development saw a drop in confidence by about 4\%. This can be attributed to the lack of back-end for the project, forcing back-end developers to learn and work in front-end development instead. Moreover, analyzing open responses showed that participants gained great insight into the development process of a software project. These include version control, CI/CD, the usage of Git (particularly pull requests), and code reviews. Participants who worked on data-related areas of the project grew 40\% more confident in data visualization and 30\% more confident in data collection. No participants indicated their ability to provide data updates had declined, with confidence in that area growing by roughly 27\%. It is worth noting that out of 10 participants who worked in promotion but not data-related areas, 8 participants did not recognize growth in their data collection and visualization skills. In addition, other technical areas where the participants felt more confident were language translation (36\%), UI/UX (32\%), and promotion (30\%).

\subsubsection{Non-Technical Skills}\label{sec:softskills} 

\begin{figure}[h]
    \centering
    \includegraphics[width=8.5cm,height=6cm,keepaspectratio]{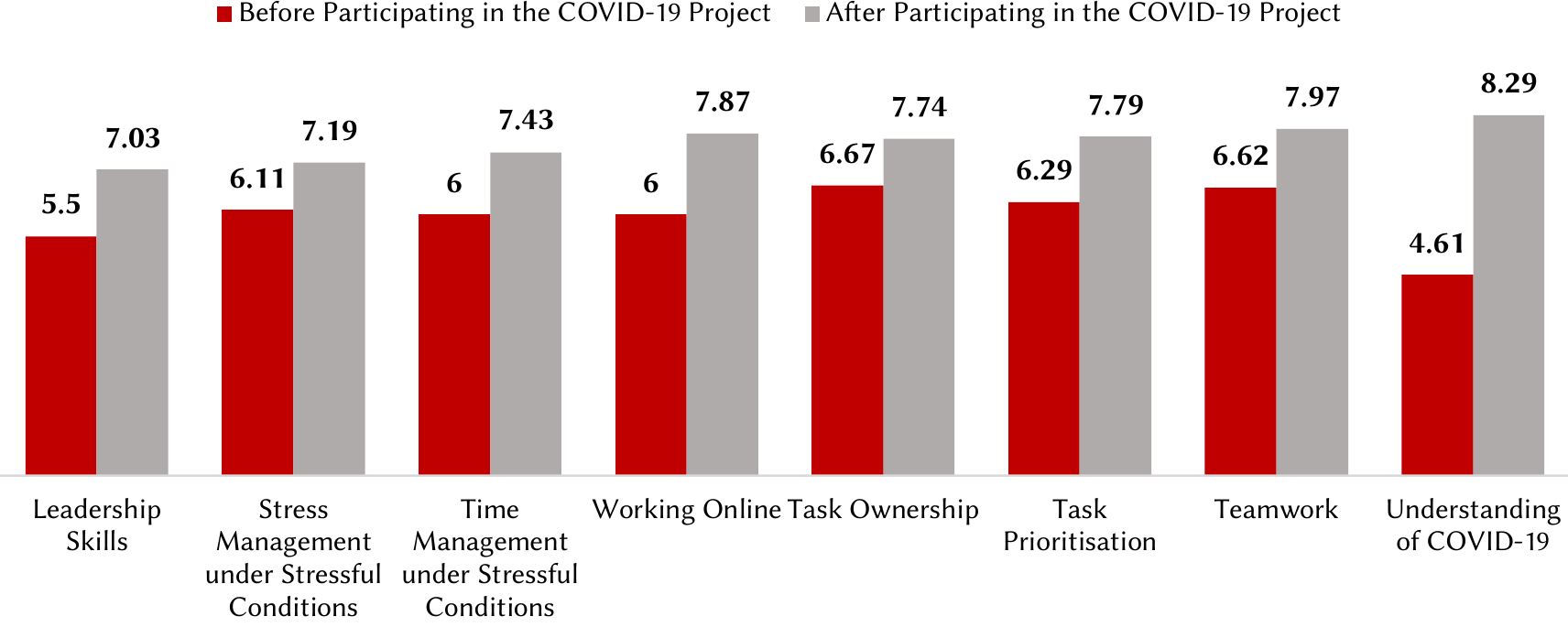}
    \caption{The mean of the participants' confidence in non-technical skills before and after participating in the COVID-19 project }
    \label{fig:sofskills}
\end{figure}

Although the participants were abruptly forced to adjust to the unusual circumstances of the pandemic, they showed high levels of learning and adaptability. Fig. \ref{fig:sofskills} indicates an overwhelming consensus among the participants that their understanding of COVID-19 grew by close to 80\% through their experience working on the project. They also grew roughly 31\% more confident in their ability to work online, with one student explicitly mentioning Slack as a useful tool for online work. Moreover, their confidence in time and stress management under stressful conditions grew by 24\% and 18\% respectively, with no participants indicating they had regressed in their ability to manage these aspects even under the challenging circumstances. 

All contributors expressed they had grown roughly 20\% more confident in teamwork. About 20 participants who stepped up to lead in specific areas or in discussions reported they grew about 28\% more confident in their leadership skills. Students especially noticed a rise in self-confidence, expressing they felt more confident \emph{``speaking up''} or \emph{``talking a lot with sponsors''}. One student identified networking as an area of growth, saying the project was a \emph{``great place to meet people''} and he \emph{``made some connections [he] was able to use''} going forward. Participants also reported they grew 24\% more confident in their ability to prioritize tasks effectively and 16\% more confident in task ownership, both of which are skills related to workload management.

\begin{center}
\begin{tcolorbox}[colback=black!5!white,colframe=black!75!black]
\textbf{\footnotesize RQ4.} \textit{Students showed overall positive growth in technical and non-technical skills, particularly data-related skills, understanding of the pandemic, and working online. Students especially appreciated learning about project development aspects such as version control and CI/CD. Back-end development was the only skill participants did not grow in due to the absence of a back-end for the project.}
\end{tcolorbox}
\end{center}

\section{Discussion}\label{sec:Discussion} 
\subsection{Implications for Research}
\subsubsection{Student Motivation}
Silva et al. found that students were motivated to participate in OSS projects by rewards and for experience \cite{silva2020google}. Our study supports this with a majority of students expressing interest in learning new skills, gaining experience for their CV, and building a network. Additionally, we observed that students were motivated to contribute by wanting to play a positive role during the pandemic and to gain experience working in a crisis. This further extends prior research as it provides an altruistic view for student motivation. Future projects may want to leverage students as reliable and responsible human resources in dealing with crises.

\subsubsection{Challenges and Growth}
Prior studies on professional developers identified several challenges stemming from working from home during the pandemic, such as a poor work environment, poor work-life balance, and difficulties in collaborating \cite{ford2020taleNew, bao2020does}. While these issues were also present in the student team, they were reported to a less significant degree. We also observed that the students had to balance their academic commitments, similar to professional developers balancing external commitments such as family \cite{ford2020taleNew, ralph2020pandemic}. Students who participate in such projects may be more adept at handling multiple responsibilities when entering the professional environment, which is supported by increased confidence in their task prioritization abilities. The main challenge of the COVID-19 project was data collection. Our findings suggest that the students were able to adapt to these difficulties through the employment of various data collection strategies. This resulted in students' growth in their data collection skills and contextual understanding of COVID-19. Future projects should encourage placing students in challenging environments as they may stimulate an improvement in skills.

\subsubsection{Student Well-being}
A major focus has been placed on developer well-being during COVID-19 \cite{ralph2020pandemic, butler2020challenges}, as WFH has often been observed to have varying impacts on an individuals well-being. The students in this project were active in maintaining the well-being of the team, with students supporting and acknowledging each other to boost team morale. This is supported by a majority expressing that mental health was not a significant challenge while working on this project. The strategies adopted by the students in maintaining well-being while working under these situations may be translated to other student teams or professional teams.

\subsection{Implications for Practice}
\subsubsection{Students}
We observed that the students employed many strategies and practices while working on this project, and conversely, an overall disagreement with the presented challenge statements. Students who work on future projects can actively utilize the strategies discussed in this study to manage their project or team, especially if working online or within a crisis.

\subsubsection{Educators}
Proper management of the challenges in a project will affect the experiences and continual motivation of the students contributing to a project. Educators can leverage the challenges identified in this study, with the corresponding strategies, to help cultivate positive experiences while working. Additionally, educators can emphasize the benefits of working on a project in line with the motivations highlighted, such as gaining experience for CV.

\subsubsection{Government}
As discussed, the students primarily faced challenges with collecting data, as the dissemination of information was disparate and inconsistent. As the primary source of information for the students was government reporting, we recommend that efforts should be made to ensure the consistency and reliability of federal reporting across Australia. This may include presenting information in a consistent format or having a central information hub. This would be beneficial not solely to data aggregation projects, but to the general public seeking information on a crisis. 

\subsection{Threats to Validity} 
\subsubsection{Threats to External Validity}
Threats to external validity refer to how our results may be generalized to other software teams of a similar nature \cite{wohlin2012experimentation}. In this study, we examined only one project team in the context of the COVID-19 pandemic. Different student teams may have different experiences depending on the nature of the students or the type of software being developed. Our findings may not generalize to other student software projects. However, the proposed methodology may provide a guideline for further studies on a larger scale.
\subsubsection{Threats to Internal Validity}
Threats to internal validity refer to the conditions present while conducting this study. We primarily ascertained the participants' opinions through Likert scales. The statements presented may have been biased based on preconceptions of the author. However, these statements were derived from a pilot study conducted on communication between the team members and existing literature to reduce this bias. Furthermore, the statements were finalized after several iterations and taking into account the opinions of multiple authors. This helped further reduce bias during the pilot study. We paired open-ended questions with each Likert scale question to allow the participants to express further views above the statements presented. While their responses could have been influenced by the preceding statements, we emphasized that further views should be separate from the presented statements.

A monetary incentive was also provided to the participants to encourage a higher quality of response. We do not present role-specific statements aside from data collection, as the pilot study indicated that there were few role-specific challenges present in other roles.
The students may have tended to give socially acceptable responses (e.g., social desirability bias \cite{furnham1986response}). We minimized this bias by assuring the students that we would not reveal their individual information and responses in any possible research outcomes.

\section{Related Work}\label{sec:RelatedWork} 

\subsection{Software Development during the COVID-19 Pandemic}
The COVID-19 pandemic has forced many organizations to adopt the work from home (WFH) policy, requiring employees to work remotely from home. Several studies, such as \cite{perez2002benefits, ford2019remote, baker2007satisfaction}, investigated the possible positive and negative impacts of WFH on employees (e.g., increased flexibility and autonomy, reduced costs) before the COVID-19 pandemic. The majority of them targeted general workers, not only software practitioners \cite{bao2020does}. 

Further to this, WFH during the COVID-19 pandemic is different from traditional remote working \cite{o2021covid, rodeghero2020please, ralph2020pandemic}. Several reasons can explain this difference, for example, having additional restrictions (e.g., restricted traveling) and blended family life and work-life during the pandemic \cite{o2021covid}. Hence, some studies have recently investigated how WFH during the pandemic has changed the way developers work and its effects on developers' productivity levels and well-being. Butler et al. \cite{butler2020challenges} found that while this policy is associated with increased flexibility and spending more time with family, software engineers in a large organization often complain about the increased workload and have too many meetings due to this policy. Through a two-wave longitudinal study, Russo et al. \cite{russo2021daily} found that WFH does not \textit{per se} presents a difficulty for either organizations or software engineers. Rodeghero et al. \cite{rodeghero2020please} looked at the remote onboarding process during the pandemic. They found that newly hired software engineers faced several challenges, such as searching internal documents, constructing team connections, and seeking help and feedback. 

Ford et al. \cite{ford2020taleNew}, Forsgen \cite{octoverse}, and Russo et al. \cite{russo2021predictors} identified lower initial productivity of developers in the early stages of the pandemic, which eventually stabilized or recovered as developers became accustomed to the new policy in place. The studies done by Ford et al. \cite{ford2020taleNew}, Bao et al. \cite{bao2020does}, and Ralph et al. \cite{ralph2020pandemic} revealed that developer productivity is affected by their adaptability to the change in work environment and ergonomics. In these studies, developers with increased productivity quoted benefits of WFH, such as better work-life balance and flexibility in working hours. Conversely, flexibility in working hours also hindered productivity when it is not well-managed by developers \cite{ford2020taleNew}. Other reasons for lowered productivity quoted were difficulties collaborating with others \cite{bao2020does}, poor work environment \cite{bezerra2020human}, distractions \cite{ford2020taleNew, bezerra2020human}, and emotional issues \cite{bezerra2020human}. 

WFH also affected developers' well-being both positively and negatively. Ralph et al. \cite{ralph2020pandemic} and Russo et al. \cite{russo2021predictors} explored the relationship between developers' productivity and well-being during the pandemic and found that they are related. Developers whose well-being improved commonly mentioned better work-life balance. Varying reasons, such as less physical activities, less social interactions, and difficult communication, were quoted by developers suffering from worse well-being \cite{ford2020taleNew}. Longer workdays and working during weekends were also observed by Forsgen \cite{octoverse}. However, the WFH policy has encouraged more collaboration on OSS, evident from the increase in the number of developers in numerous projects.

As these studies examined professional developers, ours provides insights into how students fare in a different work environment and its impact on their software development.

\subsection{Education during the COVID-19 Pandemic}
Temporal closures on most educational institutions worldwide due to the COVID-19 pandemic \cite{unesco2020education} had a tremendous impact on the education system in various countries \cite{schleicher2020impact}. In particular, schools transitioned to remote learning to continue with the curriculum. Bao \cite{bao2020covid} looked into challenges and provided suggestions (e.g., the teaching content should be divided into smaller units to help students focus) to improve the overall experience of remote learning. It was realized that students generally had lower motivation and engagement levels with remote learning. Gonzalez et al. investigated how well university students coped during the pandemic in terms of performance and found that students developed better studying habits during this time \cite{gonzalez2020influence}. Other researchers \cite{zhang2020web, kanij2020adapting, barr2020online} investigated teaching challenges and adaptive delivery methods for software engineering courses. Zhang et al. \cite{zhang2020web} found it is possible to successfully run the distributed software development course as a joint course in different countries during the pandemic by leveraging only online platforms. 


\subsection{Students in OSS Development}
Some studies have leveraged OSS in software engineering education and investigated the involvement of students in OSS development \cite{silva2020google, pinto2017training, pinto2019training, holmes2014lessons,smith2014selecting}. Silva et al. \cite{silva2020google} looked at the students' motivations in joining an OSS program and found that rewards and gaining experiences were the main motivating factors. In two studies \cite{pinto2019training, pinto2017training}, it was found that using OSS in software engineering courses is associated with benefits, such as improving technical and non-technical skills, for students and costs/challenges, such as selecting the best-fitting OSS project, for lecturers and students. Holmes et al. \cite{holmes2014lessons} argued that a mentor should be assigned to help students when they are involved in an OSS project. None of these studies explored the experiences of a student team in developing an OSS project during a pandemic. Further to this, unlike the works mentioned above, the student team that we studied had no mentor, and they had to self-manage and develop the necessary policies and processes for project development (See Section \ref{sec:developmentprocess}).
\section{Conclusions}\label{sec:Conclusion}

We conducted an online survey to understand the experiences of the students who contributed to a COVID-19 information dashboard project during the COVID-19 pandemic.
We have observed that playing a positive role in the COVID-19 crisis and learning new skills and technologies were among the most cited motivating factors for the students. The students felt that collecting data from government sources was challenging due to the inaccurate and unreliable nature of the available data. Conversely, while some students found it difficult to manage their work-life balance or form relationships with the team members, they were generally able to adapt well to the challenges while being socially isolated. Some of the strategies they used to adapt to or minimize the present challenges were frequent and clear communication in the team and uplifting team members with words of encouragement. Finally, the participants indicated an overall positive growth in their skills - especially data-related skills, understanding of the COVID-19 pandemic, working online, and an overall increase in their confidence. We hope that our research provides a clearer, deeper picture of the student response and experience in crises and contributes to identifying appropriate challenges and solutions in crisis-based software development.



\bibliographystyle{IEEEtran}
\bibliography{main}

\end{document}